\documentclass[preprint2]{aastex}
\shorttitle{Dynamics of the dE FS76}
\shortauthors{De Rijcke {\em et al.}}

\begin{document}

\title{The dynamics of the dE galaxy FS76~: bridging the kinematic dichotomy
between Es and dEs. \altaffilmark{1}}

\author{S. De Rijcke \altaffilmark{2,3} and H. Dejonghe,}
\affil{Sterrenkundig Observatorium, Universiteit Gent, Krijgslaan 281,
  S9, B-9000 Gent, Belgium}
\email{sven.derijcke@rug.ac.be}
\email{herwig.dejonghe@rug.ac.be}
\author{W.~W.  Zeilinger}
\affil{Institut f\"ur Astronomie, Universit\"at Wien,
  T\"urkenschanzstra{\ss}e 17, A-1180 Wien, Austria}
\email{zeilinger@astro.univie.ac.at}
\and
\author{G.~K.~T. Hau}
\affil{Departamento
  de Astronom\'{\i}a y Astrof\'{\i}sica, Pontificia Universidad Cat\'olica de
  Chile, Campus San Joaqu\'{\i}n, \\ Vicu\~na Mackenna 4860, Casilla 306,
  Santiago 22, Chile}
\email{ghau@astro.puc.cl}
\altaffiltext{1}{Based on observations collected at the
    European Southern Observatory, Chile (ESO Large Programme
    Nr.~165.N-0115).}
\altaffiltext{2}{Postdoctoral Fellow of the Fund for
  Scientific Research - Flanders (Belgium)(F.W.O)}
\altaffiltext{3}{Present address~: Astronomisches Institut,
  Universit\"at Basel, Venusstrasse 7, 4102 Binningen, Switzerland}

\begin{abstract}
  We present major and minor axis kinematics out to $\approx 2$
  half-light radii for the bright ($M_B=-16.7$) dwarf elliptical (dE)
  FS76, a member of the NGC5044 group. Its velocity dispersion is $46
  \pm 2$~km/s in the center and rises to $ 70 \pm 10$~km/s at
  half-light radius. Beyond $1~R_{\rm e}$ the dispersion starts to
  fall again. The maximum rotation velocity is $15 \pm 6$~km/s, about
  the value expected for an oblate isotropic rotator with the same
  flattening as FS76 (i.e. E1). Hence, FS76 is the first dE discovered
  so far that is not flattened predominantly by anisotropy. There is a
  discontinuity in the radial velocity profile at $\pm 1''$,
  corresponding to a kinematically peculiar core with a radial extent
  of 0.25~kpc. The reversed outward trend of the velocity dispersion
  is interpreted as evidence for a truncated dark halo and hence for
  the occurrence of tidal stripping. Using dynamical models we estimate
  the total mass within a sphere of 1~kpc ($\approx 1.5 R_{\rm e}$) to
  be between 1.2 and 3.4~$\times 10^9 M_\odot$ at the 90\% confidence
  level, corresponding to $3.2 \le \left( \frac{M}{L} \right)_B \le
  9.1$. These values are consistent with predictions based on CDM
  cosmological scenarios for galaxy formation.
\end{abstract}


\section{Introduction}

Dwarf ellipticals are of particular interest because they dominate in
numbers the nearby universe and in CDM scenarios they are supposed to
be dark matter dominated systems. Due to their very faint surface
brightness levels that make long slit spectroscopy a very daunting
task, the kinematics of dEs have so far been left largely unexplored.
Up to now, the central velocity dispersions of only a few tens of dEs
have been determined \citep{pc}. Furthermore, the projected velocity
and velocity dispersion profiles of only 6 dEs can be found in the
literature. These are the Virgo dEs VCC351 and IC794, the three dE
companions of M31~: NGC205, NGC185 and NGC147 and the Fornax dE
\citep{bn,ma,ba,cs,ha1,ha2}. None of the dEs in this sample are
flattened by rotation. The ratio of the observed $v/\sigma$ to the
theoretical estimate for an oblate, isotropic rotator $(v/\sigma)^* =
(v/\sigma)_{\rm obs} /(v/\sigma)_{\rm theo}$ gives an indication of
the relative importance of rotation in flattening a galaxy. All dEs in
this sample have $(v/\sigma)^*<0.4$ and hence are supported by
anisotropy. This finding introduced a dichotomy in the otherwise
linear sequence of increasing rotational support with decreasing
luminosity for ellipticals. The issue of a photometric dichotomy
between Es and dEs however has been resolved. It is now known that the
surface brightness profiles of both Es and dEs can be well
approximated by a S\'ersic profile $I(r) =I_0 \exp(-(r/r_0)^{1/n})$
and that both types form a single sequence in $(n,M_B)$, $(I_0,M_B)$
and $(r_0,M_B)$ diagrams \citep{jb}. A comprehensive review on the
subject is given by \citet{fb}. Recently \citet{rtpl} pointed out that
dEs have the same range and frequency of boxy and disky isophotes as
normal Es.

The class of dEs is a key factor in understanding galaxy formation. In
CDM cosmology scenarios, dEs form from average-amplitude density
fluctuations \citep{ds}. Supernova-driven winds are thought
responsible for expelling their gas and reshaping the galaxies. This
scenario successfully predicts the observed scaling relations between
mass, velocity dispersion and luminosity but fails to reproduce the
observed clustering properties of dEs. Alternatively, \citet{moo}
argue that dEs can also form out of late-type galaxies that are
stripped of their disk material and dark matter by galaxy harassment.
Their simulations yield objects that are qualitatively similar to
bright, non-nucleated dEs. The latter are known to have the same
clustering properties as late-type spirals \citep{fb}, strengthening
the idea of an evolutionary link between both types of objects. A
better knowledge on the internal dynamical structure of dEs is
urgently needed to test these formation scenarios.  In a case study on
the dE FS76, first results of an ongoing ESO {\em Large Programme} on
physical properties of a sample of dwarfs ranging from dE0 to dS0 are
reported. FS76 is a member of the NGC 5044 group. VRI photometry and
deep, high resolution major and minor axes kinematics are presented.
These data show that FS76 is the fastest rotating dE recorded so far
and bridges the kinematic dichotomy between Es and dEs. We argue that
its outward declining velocity dispersion is the result of a compact
dark halo.

\section{Observations and data reduction} \label{obsred}

The observations were carried out in the period May 9-16, 2000 at the
8.2~m telescope Kueyen ({\tt VLT}-{\tt UT2}) using {\tt FORS2}, both
for the imaging and the spectroscopy. It is classified as a round dE,
possibly a compact elliptical (cE). Major (PA=46$^\circ$) and minor
axis (PA=136$^\circ$) long-slit spectra were obtained with the {\tt
  FORS2} grism {\tt GRIS\_1028z+29} in the wavelength region
$\lambda\lambda 7900 -9300${\AA} achieving an instrumental broadening
of $\sigma_{\rm instr} = 30$~km/s. Total integration time was 5~h for
each position angle. The spectra were obtained at typical seeing
conditions of $0.7''-0.8''$~FWHM. The standard data reduction
procedures were performed with {\tt ESO-MIDAS} \footnote{{\tt
    ESO-MIDAS} is developed and maintained by the European Southern
  Observatory} (the details of the observations and data reduction
will be given in a subsequent paper (Dejonghe {\em et al.}  2001)).
All spectra were rebinned to a linear wavelength scale (rectifying the
emission lines of the arc spectra to an accuracy of
$\approx$1~km/s~FWHM). We also obtained spectra of 9 giant stars in
the late G to early M spectral range as velocity templates. In
addition, we obtained V, R and I images during a period of excellent
seeing ($\approx 0.3''-0.4''$~FWHM).

\begin{figure}
\plotone{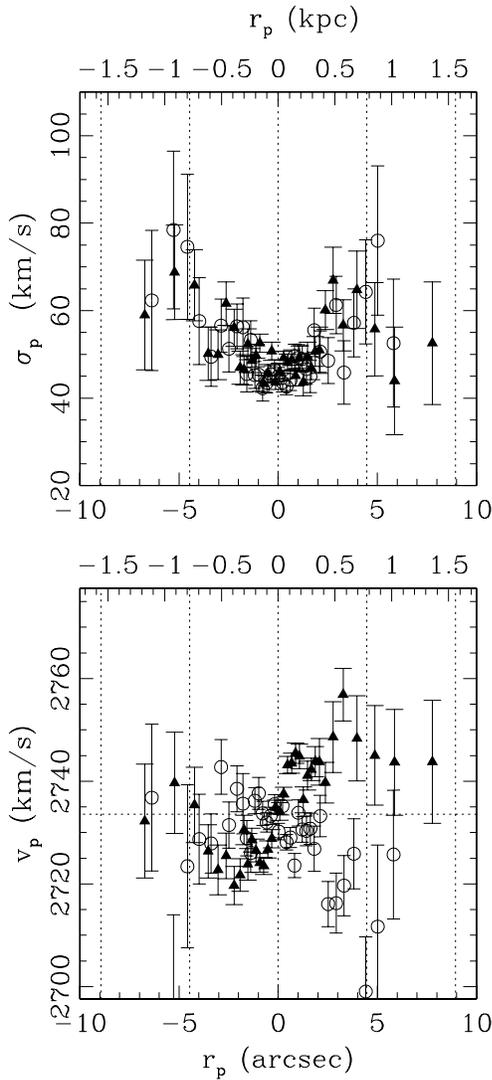}
\caption{The major (triangles) and minor axis (circles) kinematics of FS76. 
  Lower panel~: the mean projected velocity $v_p$, top panel~: the
  projected velocity dispersion $\sigma_p$. The linear scale (in kpc)
  is shown at the top axis. The vertical dashed lines measure
  distances of one and two half-light radii. The horizontal dashed
  line in the lower panel marks the systemic velocity of FS76.
  \label{kinps}}
\end{figure}

Ellipses were fitted to the isophotes of the calibrated VRI images in
order to derive surface brightness, position angle and ellipticity
profiles together with the Fourier coefficients that quantify the
deviations of the isophotes from a pure elliptic shape. Errors on all
quantities were estimated using the bootstrap method. S\'ersic
profiles were fitted to the V, R and I band growth curves. The derived
photometric characteristics of FS76 can be found in Tables
\ref{photcoloro} and \ref{sersic}. The kinematic parameters -- the
mean rotation velocity $v_p$ and the velocity dispersion $\sigma_p$ --
were obtained by fitting Gaussians to the line-of-sight velocity
distributions (LOSVDs). The best fit was obtained with a K1{\sc iii}
template. Template mismatch was found to be negligible with this star
spectrum. The resulting mean velocity and velocity dispersion
profiles for major and minor axes are presented in Figure \ref{kinps}.

\section{Results and discussion}

The analysis of the surface photometry confirms the picture of FS76
being a normal dwarf elliptical. No photometric peculiarities were
noted. There is only a modest amount of isophote twisting (the PA
varies slowly between $30^\circ$ and 50$^\circ$ over a radial region
of $10''$). FS76 is not nucleated and its nucleus (defined as the
brightest pixel) is coincident with the geometric center of the outer
isophotes. Work on the Virgo cluster by \citet{sbt}, has shown that
only 20\% of all dEs of comparable intrinsic brightness are
non-nucleated. No significant deviations from ellipses were detected
in the isophotes. A heliocentric velocity of $2734 \pm 4$ km/s was
derived from the spectra which confirms FS76 as a member of the
NGC5044 group ($v_{\rm NGC5044} = 2704 \pm 33$~km/s). Using a distance
modulus of $-32.78$ ($H_0 = 75$~km/s/Mpc) for the NGC5044 group we
derive $M_B=-16.7$~mag for FS76. The galaxy closely follows the
relations between the S\'ersic parameters and absolute magnitude found
to be valid for Es and dEs (but not for cEs) and also falls among the
bright dEs in a central surface brightness versus absolute magnitude
diagram with $m_B(0) = 19.15$~mag. We can therefore positively exclude
a cE-type nature of FS76 and classify it as dE1.

The velocity dispersion profiles of both axes agree remarkably well. A
central value of $\sigma_{\rm p} = 46 \pm 2$ km/s is derived. The
velocity dispersion rises outward to about $70 \pm 10$ km/s at
half-light radius ($R_{\rm e}$). Beyond $1 R_{\rm e}$ the profile
declines to about 50~km/s at a radial distance of $2 R_{\rm e}$. The
maximum rotation velocity along the major axis is $15 \pm 6$ km/s.  A
discontinuity in the radial velocity profile at $\pm 1''$, indicates
the presence of a kinematically decoupled core. The rotation velocity
profile shows a distinct asymmetry. There is a hint of this asymmetry
in the velocity dispersion profile as well. A small velocity gradient
is also present along the minor axis. This immediately excludes
another alternative, i.e. that we are seeing a disk galaxy almost
face-on, because such an object would have a much lower surface
brightness, would lack minor-axis rotation and have a much smaller
velocity dispersion.

The ratio of the rotation velocity to the velocity dispersion can be
used as an indicator for the importance of rotation in flattening a
stellar system. Using equation (4-95) of \citet{bt}, the expected
ratio for an isotropic E1 galaxy is $v_p/\sigma_p= 0.41$. Another
theoretical estimate yields $v_p/\sigma_p = \pi \sqrt{
  2((1-\epsilon)^{-0.9}-1)} /4 = 0.35$ (e.g. \citet{sg}). The observed
ratio of the peak velocity to the central velocity dispersion is
$(v_p/\sigma_p)_{\rm obs} = 0.33 \pm 0.15$, corresponding to an
anisotropy parameter $(v/\sigma)^* \approx 0.9$, the highest value
obtained so far for a dE. Moreover, the best fitting dynamical model
(see section \ref{model}) learns that the radial pressure drops only
very little towards the rotation axis (the radial velocity dispersion
$\sigma_r$ varies only by a few km/s if one moves from the equatorial
plane towards the rotation axis) and therefore pressure differences
play only a minor role in flattening this galaxy. Thus, the observed
kinematics and detailed dynamical models unambiguously show that FS76
is indeed flattened by rotation and not by pressure anisotropy. This
suggests that at least in some dEs rotation plays a major role. Of the
6 galaxies in our present sample, which includes also very flattened
objects, of which the kinematics have been analyzed \citep{d2001},
this is the only one that is rotationally flattened. Hence, if we add
also the dEs for which spatially resolved kinematics have been
published and interpret small-number statistics, the current sample
of dEs suggests that probably not more than 10\% of all dEs are
flattened by rotation.

\section{The dynamical modeling} \label{model}

The derived kinematics together with the reported results from the
surface photometry argue in favor of an almost round though slightly
triaxial object, viewed approximately along its intermediate axis. For
the modeling purposes we assumed a spherical gravitational potential
with the stellar body seen edge-on. The total spatial density was
taken to be the spatial density of the luminous matter (obtained by
deprojecting the surface brightness profile), multiplied with a
constant or outwardly rising function, corresponding respectively to a
constant or outwardly rising $M/L$. The gravitational potential
follows from Poisson's equation. The orbital structure is described by
the distribution function (DF) $F(E,L,L_z)$, a function of binding
energy $E$, angular momentum $L$ and the component of the angular
momentum along the rotation axis $L_z$. This DF can be obtained by
fitting a dynamical model directly to the spectra \citep{dj}. Thus,
all kinematic information in the spectra is used, the shape of the
LOSVD is not biased towards a Gaussian and each model can be given an
absolute likelihood (the $\chi^2$ serves as a goodness of fit). By
trying a wide variety of mass distributions, the range of models that
are compatible to the data can be determined, yielding reliable
$M/L$-estimates.

\begin{figure}
\plotone{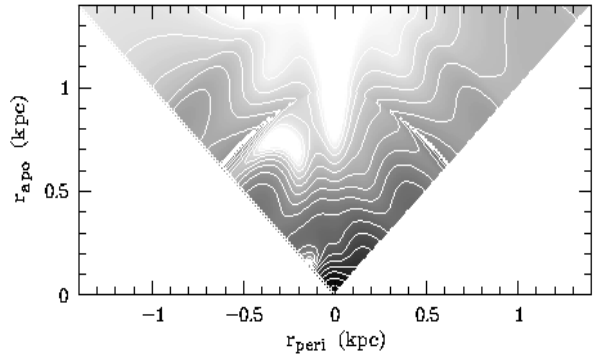}
\caption{The DF of FS76 for equatorial plane orbits in turning-point space, i.e. the 
  phase-space density of stars on orbits with a given pericenter
  distance $r_{\rm peri}$ and apocenter distance $r_{\rm apo}$. The
  contours are spaced 0.25 logarithmic bins apart. $r_{\rm peri}$ has
  the same sign as the angular momentum, $L_z$. The peculiar core
  consists of stars around the locus of circular orbits with radius
  $\approx 0.2$~kpc ($\approx 1''$). Mark also the paucity of stars on
  radial orbits with apocenter distances larger than 0.7~kpc ($4''$).
  At all radii, there are more stars on orbits with positive $L_z$
  than with negative $L_z$, producing the bulk rotation of the galaxy.
  \label{dfps}}
\end{figure}

We fitted about 100 models to the spectra, with mass distributions
varying between constant $M/L$ and models with steeply outwardly
rising $M/L$ (consisting of up to 50\% of dark matter inside 1~kpc).
The total mass within a sphere of 1~kpc ($\approx 1.5 R_{\rm e}$) is
found to be between 1.2 and 3.4~$\times 10^9 M_\odot$ at the 90\%
confidence level. This corresponds to $3.2 \le \left( M/L \right)_B
\le 9.1$. Although a model with constant $(M/L)_B=5.6$ cannot be ruled
out, the best fitting model has an outwardly rising $M/L$ ($\left( M/L
\right)_B = 6.0$ at 1~kpc). This model remains isotropic out to
0.25~kpc and becomes slightly tangentially anisotropic at larger radii
(the radial dispersion is about 40~km/s while the tangential
dispersion is 65~km/s). Analysis of the distribution function revealed
that the kinematically decoupled core is indeed due to a fast
rotating, isotropic stellar component with a radial extent of 0.25~kpc
(contrary to the case of NGC7097 where analysis of the distribution
function indicated that the counter-rotation is not related to a
compact group of stars \citep{dbr}). The DF for orbits in the
equatorial plane is plotted in Figure \ref{dfps}.

\citet{ds} have derived scaling relations between the luminosity $L$,
radius $R$ and total mass $M$ of elliptical galaxies. Their simple
model of galaxy evolution is based on two premises~: ($i$) dEs lose
significant amounts of gas in a supernova-driven wind and are
surrounded by a dark halo that dominates their dynamics and ($ii$) the
observed relation $L \propto R^4$ between luminosity and radius holds
for all ellipticals. They predict that fainter dEs are more dark
matter dominated ($M/L \propto L^{-0.37}$) and have smaller velocity
dispersions ($\sigma \propto L^{0.19}$). The derived relation between
mass and radius ($M \propto R^{2.5}$) happens to correspond to what is
predicted by CDM cosmological models. Hence, obtaining reliable $M/L$
estimates for dEs has important repercussions on cosmology. FS76
closely follows the $L \propto R^4$ and $\sigma \propto L^{0.19}$
sequences (we're using the central velocity dispersion for
comparison). Dekel~\&~Silk predict $(M/L)_B \approx 3$, somewhat lower
than what we find. This is likely the result of the outwardly rising
velocity dispersion profile. Overall, we can say that our results for
FS76 are in reasonable agreement with the predictions of the standard
CDM cosmological scenarios for galaxy formation.

\section{The compact halo of FS76}

A peculiarity of FS76 is the reversed outward trend of its velocity
dispersion~: it rises from 46~km/s in the center to about 70~km/s at
half-light radius and then declines to approximately 50~km/s at 2
half-light radii. This feature is present on both sides of major axis
and minor axis. If radial anisotropy is responsible for the falling
velocity dispersion at large radii, one would see a high velocity
dispersion in the center since there the line of sight is almost
parallel to the radial orbits, something which is not observed.
Moreover, our dynamical models show no sign of a rapidly changing
anisotropy at the radius where the reversed trend sets in~: they
remain slightly tangentially anisotropic. We also find that dynamical
models with rapidly rising $M/L$ have too high a velocity dispersion
at large radii. Therefore we consider this to be strong evidence that
the halo, which according to our modeling is likely to be present,
must be compact and thus of comparable extent as the luminous matter,
indicating that, at least for this dE, the light is not `the top of
the iceberg'. This would fit into the scenario suggested by
\citet{moo} where dark matter dominated disk galaxies are transformed
into baryon dominated dEs. The kinematically peculiar core we observe
may be a consequence of gas that is driven towards the center by the
torques induced by tidal interactions, followed by star-formation.
Moreover, FS76 is a bright non-nucleated dE. This species of dEs is
clustered the same way as spirals, which led \citet{fb} to suggest a
possible evolutionary link between these dEs and late-type galaxies.
An $M/L$ in the range 3 to 8 is predicted from this scenario,
consistent with our findings.

\acknowledgments WWZ acknowledges the support of the Austrian Science
Fund (project P14783) and of the Bundesministerium f\"ur Bildung,
Wissenschaft und Kultur. GKTH acknowledges support of Chilean FONDECYT
grant no. 1990442. SDR acknowledges financial support of the Belgian
Fund for Scientific Research. SDR also would like to thank Bruno
Binggeli for interesting discussions.

\clearpage

\clearpage

\begin{deluxetable}{ccccccc}
\tabletypesize{\scriptsize}
\tablecaption{Extinction corrected (m$_{\rm T}^0$) apparent 
  magnitudes and colors, effective surface brightnesses $\langle \mu
  \rangle _{\rm e}^0$ and half-light radii $R_{\rm e}$ of FS76.
  \label{photcoloro}}
\tablewidth{0pt}
\tablehead{
\colhead{} & \colhead{V}  & \colhead{R}   &   \colhead{I}    & \colhead{V$-$R} & \colhead{V$-$I} 
& \colhead{R$-$I}  }
\startdata
  m$_{\rm T}^0$         & 15.44 & 14.84 & 14.35  &  0.60 & 1.09  & 0.49  \\
  $\langle \mu \rangle _{\rm e}^0$ & 20.60 & 20.03 & 19.60 & &  & \\ 
  $R_{\rm e}~('')$   & 4.29 & 4.35 & 4.47 & &  &  \\ \cline{1-7}
\enddata
\end{deluxetable}

\clearpage

\begin{deluxetable}{ccccc}
  \tabletypesize{\scriptsize} \tablecaption{Parameters of the V, R and
    I band S\'ersic profiles~: the extrapolated central surface
    brightness $m_0$, the scale-length $r_0$ and the exponent $n$.
    \label{sersic}} \tablewidth{0pt} \tablehead{ \colhead{} & \colhead{$m_0$} & 
\colhead{$r_0~('')$} & \colhead{$n$} }
  \startdata
  V    &   16.63    &  0.21 &  1.97     \\
  R    &   16.05    &  0.21 &  1.98     \\
  I    &   15.63    &  0.22 & 1.98     \\
  \enddata
\end{deluxetable}


\begin{thebibliography}{}
\bibitem[Bender~\&~Nieto(1990)]{bn} Bender R.~\&~Nieto J.-L. 1990, \aap, 239, 97
\bibitem[Bender {\em et al.}(1991)]{ba} Bender R. {\em et al.} 1991, \aap, 246, 349
\bibitem[Binney~\&~Tremaine(1987)]{bt} Binney J.~\&~Tremaine S. 1987, ``Galactic Dynamics'', 
Princeton Univ. Press, Princeton, New Jersey, USA
\bibitem[Carter~\&~Sadler(1990)]{cs} Carter D.~\&~Sadler E.~M. 1990, MNRAS, 245, 12
\bibitem[De Bruyne {\em et al.}(2001)]{dbr} De Bruyne, V. {\em et al.} 2001, \apj, 546, 903
\bibitem[Dejonghe {\em et al.}(2001)]{d2001} Dejonghe, H. {\em et al.} 2001, in preparation
\bibitem[Dekel~\&~Silk(1986)]{ds} Dekel A.~\&~Silk J. 1986, \apj, 303, 39
\bibitem[De Rijcke~\&~Dejonghe(1998)]{dj} De Rijcke S.~\&~Dejonghe H. 1998, \mnras, 298, 677
\bibitem[Ferguson~\&~Sandage(1990)]{fs} Ferguson H.~C.~\&~Sandage A. 1990, \aj, 100, 1 
\bibitem[Ferguson~\&~Binggeli(1994)]{fb} Ferguson H.~C.~\&~Binggeli B., 1994, ARAA, 6, 67 
\bibitem[Held {\em et al.}(1990)]{ha1} Held E.~V. {\em et al.} 1990, \aj, 100, 415 
\bibitem[Held {\em et al.}(1992)]{ha2} Held E.~V. {\em et al.} 1992, \aj, 103, 851 
\bibitem[Jerjen~\&~Binggeli(1998)]{jb} Jerjen H.~\&~Binggeli B. 1998, \aj, 116, 2873 
\bibitem[Mateo {\em et al.}(1991)]{ma} Mateo M. {\em et al.} 1991, \aj, 102, 914
\bibitem[Moore {\em et al.}(1998)]{moo} Moore B. {\em et al.} 1998, \apj, 495, 139
\bibitem[Peterson~\&~Caldwell(1993)]{pc} Peterson R.~C.~\&~Caldwell N. 1993, \aj, 105, 1411 
\bibitem[Ryden {\em et al.}(1999)]{rtpl} Ryden, B.~S. {\em et al.} 1999, \apj, 517, 650
\bibitem[Sandage {\em et al.}(1985)]{sbt} Sandage, A. {\em et al.} 1985, \aj, 90, 1759
\bibitem[Sparke~\&~Gallagher(2000)]{sg} Sparke L. S.~\&~Gallagher J. S., III 2000, ``Galaxies in 
the Universe: an Introduction'', Cambridge Univ. Press, Cambridge, UK
\bibitem[Yoshii~\&~Arimoto(1987)]{ya} Yoshii Y.~\&~Arimoto N. 1987, \aap, 188, 13 
\end{thebibliography}
\end{document}